\documentclass[12pt]{article}
\usepackage{amssymb,graphicx}
\usepackage{epstopdf}
\usepackage{amsmath,amsfonts}
\usepackage{epsfig}

\def\d{\partial}
\def\l{\left(}
\def\r{\right)}

\newcommand{\be}{\begin{equation}}
\newcommand{\ee}{\end{equation}}
\newcommand{\bea}{\begin{eqnarray}}
\newcommand{\eea}{\end{eqnarray}}
\newcommand{\bg}{\begin{gather}}
\newcommand{\eg}{\end{gather}}
\newcommand{\bseq}{\begin{subequations}}
\newcommand{\eseq}{\end{subequations}}

\begin{document}
\begin{flushright}
\end{flushright}
\vspace{10pt}
\begin{center}
  {\LARGE \bf Can Galileons \\[0.3cm]  support Lorentzian
wormholes? } \\
\vspace{20pt}
V. A. Rubakov\\
\vspace{15pt}
\textit{
Institute for Nuclear Research of
         the Russian Academy of Sciences,\\  60th October Anniversary
  Prospect, 7a, 117312 Moscow, Russia;}\\
\vspace{7pt}
\textit{
Department of Particle Physics and Cosmology,\\
Physics Faculty, M. V. Lomonosov Moscow State University\\ Vorobjevy Gory,
119991, Moscow, Russia
}

    \end{center}
    \vspace{5pt}

\begin{abstract}
We discuss the possibility of constructing stable, static, spherically
symmetric, asymptotically flat 
Lorentzian wormhole solutions in General Relativity 
coupled to
a generalized Galileon field $\pi$. Assuming that Minkowski space-time
is obtained at $\d \pi =0$, we find that there is tension between the 
properties of the energy-momentum tensor required to support a wormhole
(violation of average null energy conditions) and stability of the Galileon
perturbations about the putative solution 
(absence of ghosts and gradient instabilities). In 
3-dimensional space-time, this tension is strong enough to rule out 
wormholes with above properties. 
In higher dimensions, including the most
physically interesting case of 4-dimensional space-time, 
wormholes, if any, must have fairly contrived shapes.

\end{abstract}

\section{Introduction and summary}

Lorenzian wormholes~\cite{book}, if existed,
would be fascinating 
objects~\cite{Morris:1988cz,Morris:1988tu,Novikov:2007zz,Shatskiy:2008us}.
In classical General Relativity, however, 
asymptotically flat Lorenzian wormholes can be supported only by matter
that violates the null energy condition,
NEC~\cite{Morris:1988tu,hoch-visser,Hochberg:1998ha}, 
while known forms of matter
do not have this property (an interesting example of a 
Lorentzian wormhole which is not
asymptotically flat is given in Ref.~\cite{Ayon-Beato:2015eca}). 
Furthermore, in most of classical 
field theory models,
NEC violation, if any, is plagued by ghosts and/or gradient instabilities.
In particular, scalar field theories with the Lagrangians involving at most
first space-time derivatives may admit NEC-violating solutions, including 
Lorentzian wormholes (see, e.g., Ref.~\cite{Arm} and references therein),
but perturbations about these backgrounds have gradient
instabilities and/or ghosts~\cite{Buniy:2006xf}.

It is known, however, that there exist scalar field theories whose
Lagrangians contain second derivative terms, and yet whose field equations 
are second 
order~\cite{Horndeski:1974wa,Fairlie:1991qe,Luty:2003vm,Nicolis:2008in,Deffayet:2010zh,Padilla:2012dx}. 
These theories, dubbed generalized Galileon models, admit classical
NEC-violating
solutions of cosmological type, which do not have obvious 
pathologies~\cite{Genesis1,Deffayet:2010qz,Kobayashi:2010cm,Qiu:2011cy} 
(for a review see, e.g., 
Ref.~\cite{ufn}), modulo a superluminality issue~\cite{superlum,superlum2}.

It is therefore natural to ask whether generalized Galileon theories admit
stable Lorentzian wormholes within General Relativity. 
It is this question that we 
address in this paper, albeit not in full generality. The qualifications are
as follows.
First, we 
specify to a subclass of generalized Galileons, in which the
Larangians have most commonly sudied 
form~\cite{Deffayet:2010qz,Kobayashi:2010cm} 
\be
L =  F(\pi, X) + K (\pi, X) \Box \pi  \; ,
\label{sep22-15-30}
\ee
where $\pi$ is the Galileon field,
$F$ and $K$ are arbitrary Lagrangian functions, and 
\[
X = \nabla_\mu \pi \nabla^\mu \pi \; , \;\;\;\;
\Box \pi =  \nabla_\mu  \nabla^\mu \pi \; .
\]
We further assume that there is  Minkowski limit, which occurs 
at 
\[
\d_\mu \pi = 0 \; .
\]
Note that assuming that $K$ is regular at $X=0$, one can set
\be
K(\pi, X=0) = 0 \; ,
\label{sep24-15-10}
\ee
since the term $K(\pi, 0) \Box \pi$ can be absorbed into $F$ upon
integration by parts. Note also that the scalar potential, if any, is contained 
in $F$, namely, $V(\pi) = - F(\pi, 0)$.

\begin{figure}[!tb]
\centerline{\includegraphics[width=0.5\textwidth,angle=-90]{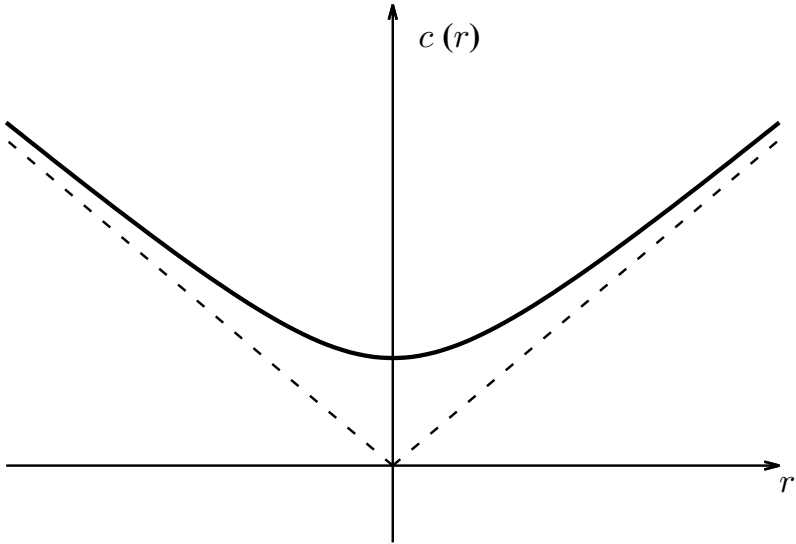}%
}
\caption{The behavior of the metric coefficient $c$.
\label{fig1}
}
\end{figure}

Second, we consider static and sphericaly symmetric wormholes
in $(d+2)$-dimensional space-time. The 
metric is  (signature $(+,-,\ldots ,-)$)
\[
ds^2 = a^2 (r) dt^2 - b^2(r) dr^2 - c^2(r) \gamma_{\alpha \beta} dx^\alpha dx^\beta
\; ,
\]
where
$x^\alpha$ and $\gamma_{\alpha \beta}$ are cordinates and
 metric on unit $d$-dimensional sphere. 
The coordinate $r$ runs from $-\infty$ to
$+\infty$, and the wormhole geometry is assumed to be asymptotically
flat. For $d \geq 2$ (four or more space-time dimensions) this implies the
asymptotic behavior
\be
d\geq 2 \; : \;\;\; a \to a_{\pm} \; , \;\;\;\; b\to 1 ,  \;\;\;\;
c(r) \to \pm r \; , \;\;\;\; \mbox{as} \;\; r\to \pm \infty \; ,
\label{sep22-15-10}
\ee
 see Fig.~\ref{fig1}, where $a_{\pm}$ are positive constants.

In 3-dimensional space-time ($d=1$) the asymptotics of $c(r)$ 
is less resricted,
\be
d = 1 \; : \;\;\; a \to a_{\pm} \; , \;\;\;\; b\to 1 ,  \;\;\;\;
c(r) \to \pm C_{\pm} r \; , \;\;\;\; \mbox{as} \;\; r\to \pm \infty \; ,
\label{sep22-15-11}
\ee
where $C_{\pm}$ are positive constants. For consistency, the 
Galileon field supposedly supporting a wormhole is also static and
spherically symmeric, $\pi = \pi (r)$, and 
\be
\pi'\to 0 \;\;\;\;\; \mbox{as} \;\; r \to \pm \infty \; ,
\label{sep23-15-2}
\ee
where prime denotes $d/d r$.

\begin{figure}[!tb]
\centerline{\includegraphics[width=0.5\textwidth,angle=-90]{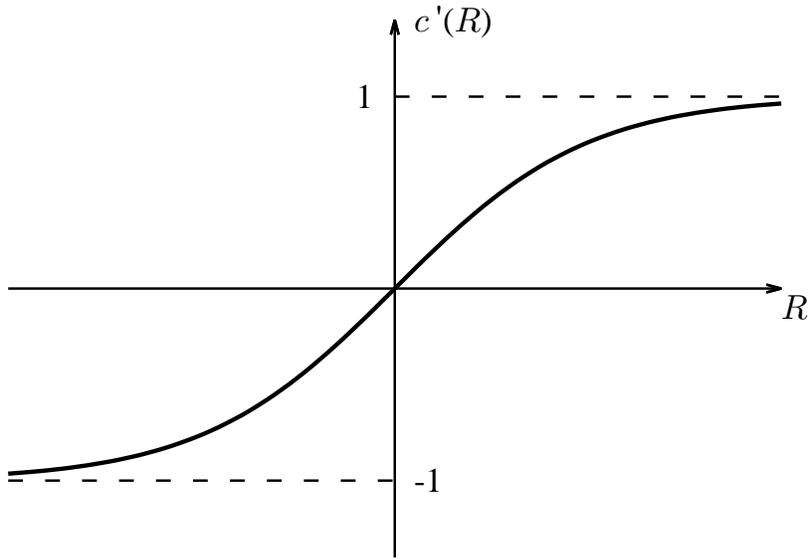}%
}
\caption{The behavior of $dc/dR$ for a wormhole of the simplest shape.
$R = \int~bdr$ is the proper radial distance.
\label{fig2}
}
\end{figure}

Our main concern is stability, namely, the absence of ghosts and 
gradient instabilities in the Galileon perturbations about the solution
$\pi(r)$. We observe that in any dimension, there is 
tension between the properties of the Galileon
energy-momentum tensor that can support a wormhole, on the one hand,
and stability requirement, on the other. In  3-dimensional space-time ($d=1$)
this tension is so strong that under very
mild assumption on the asymptotic
behavior of $\pi'$ at spatial infinity,
we show that there are no stable wormholes at all, the result 
somewhat reminiscent
of Ref.~\cite{Deser:1991ye}.
We cannot prove similar no-go theorem in 4- or higher-dimensional space-time,
but we will see that the simplest wormhole shapes are also inconsistent with
stability. By a wormhole of the simplest shape we mean a solution for
which $dc/dR$, where $R=\int b dr$ is the proper radial distance,
monotonously increases\footnote{In fact, our observation is somewhat stronger:
wormholes for which $|dc/dR| \leq 1$ at all $R$ are inconsistent with
stability.} from $-1$ to $1$ as $R$ increases from $-\infty$
to $+\infty$, see Fig.~\ref{fig2}. Similar property holds in the coordinate
frame such that\footnote{This frame
can be 
called Schwarzschild.} 
$b(r) = a^{-1}(r)$, namely,
if $dc/d r$ monotonously increases in this frame 
as $r$ increases, then the wormhole is unstable. These are 
our main results: under the assumptions stated above,
Galileons cannot support stable wormholes in 3-dimensional
space-time, while in higher dimensions, stable Galileon-supported wormholes,
at all exist, must have quite non-trivial properties.

The paper is organized as follows. We discuss the properties of the
energy-momentum tensor that supports a wormhole in Section~\ref{ANEC}.
In Section~\ref{Galileons} we consider the static, spherically symmetric
Galileons and present the form of the energy-momentum tensor and
stability conditions
of the Galileon perturbations. We derive our main results in 
Section~\ref{Tensions} and conclude in Section~\ref{Discussion}.

\section{Averaged NEC violation}
\label{ANEC}

Let us establish some features of the energy-momentum tensor
supporting a wormhole. These are similar to the properties that lead 
to averaged NEC violation
(ANEC violation), see Ref.~\cite{book} and references therein.
To this end we recall that the 
point-wise NEC violation occurs when $T_\mu^\nu k^\mu k_\nu <0$
for some null vector $k^\mu$, while the ANEC violation in broad sense is
the negaive value of some line integral of  $T_\mu^\nu k^\mu k_\nu$.
In our context the relevant null vector has components
 $k^0 = a^{-1}$, $k^r = b^{-1}$ and
$k^\alpha = 0$, so that   $T_\mu^\nu k^\mu k_\nu = T^0_0 - T^r_r$;
the line integrals we encounter
have the form $\int_{-\infty}^{+\infty}~dr \varphi (r)
\l T^0_0 - T^r_r \r$ with a positive function $\varphi (r)$.

We set $8\pi G =1$ and make use of the Einstein 
equations\footnote{$T_0^0=\rho$, $T_r^r = -p_r$, $T_\beta^\alpha =
- \delta_\beta^\alpha p_t$, where $\rho$ is the energy density, 
$p_r$ is  the radial pressure and $p_t$ is the tangential pressure.
We will not use this nomenclature in what follows.}
$T^\mu_\nu = G^\mu_\nu$, where
\begin{subequations}
\begin{align}
G_{0}^0 &= d  \left[ \frac{c' b'}{cb^3} - \frac{c^{\prime \prime}}{cb^2} -
\frac{d-1}{2}\left(\frac{c^{\prime 2}}{c^2 b^2} 
- \frac{1}{ c^2}\right) \right] \; ,
\label{sep22-15-1}\\
G_{r}^r &= - d \left[\frac{a' c'}{acb^2} + \frac{d-1}{2} \left(
\frac{c^{\prime 2}}{c^2b^2}
- \frac{1}{c^2}\right) \right] \; ,
\label{sep22-15-2}\\
G^\alpha_{\beta} &= \delta^{\alpha}_\beta G^\Omega \; ,
\label{sep2-15-3}
\end{align}
\end{subequations}
with
\begin{align}
G^\Omega = - &\left[\frac{ a^{\prime \prime}}{a b^2} - \frac{a' b'}{ab^3}
+ (d-1) \left(\frac{ c^{\prime \prime}}{cb^2} +\frac{c' a'}{cab^2}  -
\frac{c' b'}{cb^3} \right)
\right.
\nonumber \\
& + \left.
\frac{(d-1)(d-2)}{2} \left(\frac{c^{\prime 2}}{c^2b^2} - \frac{1}{c^2}\right)
\right] \; .
\nonumber
\end{align}
By combining Eqs.~\eqref{sep22-15-1} and \eqref{sep22-15-2} one obtains
\be
T^0_0 - T^r_r = -d \frac{a}{bc} \l \frac{c'}{ab} \r^\prime \; .
\label{sep22-15-15}
\ee
The latter equaltion leads to the most commonly
used form of the ANEC violation~\cite{book}, namely,
\[
\int_{-\infty}^{+\infty}~dr~ \frac{b}{a} \l T^0_0 - T^r_r \r
= - d \int_{-\infty}^{+\infty}~dr~ \frac{c^{\prime 2}}{abc^2} < 0 
\]
(the surface term appearing 
when integrating by parts
vanishes
because of the asymptotics \eqref{sep22-15-10} or
\eqref{sep22-15-11}). This can be generalized to
\be
\int_{-\infty}^{+\infty}~dr~ \frac{bc^\alpha}{a} \l T^0_0 - T^r_r \r
< 0  \;\;\;\;\; \mbox{for~all} \; \; \alpha \leq 1 \; .
\label{sep22-15-45}
\ee
The generalization is straightforward for $\alpha <1$ (the surface term again
vanishes), 
while for
$\alpha = 1$ one has
\[
\int_{-\infty}^{+\infty}~dr~ \frac{bc}{a} \l T^0_0 - T^r_r \r
= -d \l \frac{C_+}{a_+} + \frac{C_-}{a_-} \r 
\]
(where $C_{\pm} = 1$ for $d\geq 2$).

We will see that the Galileon energy-momentum tensor obeys
$T^\alpha_\beta = \delta^\alpha_\beta T^\Omega$ with
\be
  T^\Omega = T^0_0 \; ,
\label{sep22-15-31}
\ee
which implies $G^\Omega = G^0_0$ and gives
\[
- \frac{c^{\prime \prime}}{b^2 c} + \frac{c' b'}{b^3 c}
+ \frac{a^{\prime \prime}}{b^2 a} - \frac{a' b'}{b^3 a}
+ (d-1) \frac{c'a'}{b^2 ca} - (d-1)\l \frac{c^{\prime 2}}{b^2 c^2} -
\frac{1}{c^2} \r = 0 \; .
\]
We use the latter equation to cast Eq.~\eqref{sep22-15-15}
into the followinf form,
\be
T^0_0 - T^r_r = -\frac{d}{abc^{d-2}} \l \frac{a' c^{d-2}}{b} \r^\prime
- \frac{d(d-1)}{c^2}\l 1- \frac{c^{\prime 2}}{b^2} \r \; .
\ee
We will use this relation for $d\geq 2$. Assuming that
\be
  a' r^{d-2} \to 0 \;\;\;\;\; \mbox{as} \;\; r\to \pm \infty \; ,
\label{sep22-15-20}
\ee
we write
\begin{align}
\int_{-\infty}^{+\infty}~dr~abc^{d-2} \l T^0_0 - T^r_r \r 
&= - d(d-1)  \int_{-\infty}^{+\infty}~dr~abc^{d-4} 
\l 1- \frac{c^{\prime 2}}{b^2} \r 
\nonumber \\
&=  - d(d-1)  \int_{-\infty}^{+\infty}~dR~ac^{d-4} 
\left[ 1- \l \frac{dc}{dR} \r^2 \right]  
\; ,
\label{sep23-15-1}
\end{align}
where
\[
  R = \int~b~ dr
\]
is the proper radial distance.
Provided the right hand side of eq.~\eqref{sep23-15-1} 
is negative, this is another form of the
ANEC violation. Note that the assumption \eqref{sep22-15-20} is not
restictive: once the total mass seen by an outside observer
is finite,
one has Newtonian asymptotics (recall that the number of space-time dimensions
is $(d+2)$)
\[
a = 1 + O(r^{-(d-1)}) \; ,
\]
and therefore $a' = O (r^{-d})$.

\section{Static, spherically-symmetric Galileons}
\label{Galileons}

\subsection{Energy-momentum tensor}

Turning to Galileons, the energy-momentum tensor of a theory with the
Lagrangian~\eqref{sep22-15-30} reads, in general,
\be
T_{\mu \nu} = 2 F_X \d_\mu \pi \d_\nu \pi
+  2 K_X \Box \pi \cdot \d_\mu \pi \d_\nu \pi
- \d_\mu K \d_\nu \pi - \d_\nu K \d_\mu \pi - g_{\mu \nu} F
+ g_{\mu \nu} g^{\lambda \rho} \d_\lambda K \d_\rho \pi \; ,
\nonumber\\
\ee
where $F_\pi = \d F/\d \pi$, $F_X = \d F/\d X$,
etc. (we reserve prime for $d/d r$),
and $\d_\mu K = K_\pi \d_\mu \pi + 2 K_X \nabla^\lambda \pi \nabla_\mu
\nabla_\lambda \pi$. We immediately see from this expression that
in the static spherically-symmetric case at hand, when $\pi = \pi (r)$,
the energy-momentum tensor has the property \eqref{sep22-15-31}.
We have, explicitly,
\begin{align}
T_{0}^0 & =  -F  - K_\pi \left(\frac{\pi^\prime}{b} \right)^2
+ 2  \left(\frac{\pi'}{b} \right)^2  
\frac{1}{b}\left(\frac{\pi'}{b} \right)^\prime K_X  \; ,
\nonumber\\
T_{r}^r &= - 2  \left(\frac{\pi'}{b} \right)^2 F_X -F 
 + K_\pi \left(\frac{\pi^\prime}{b} \right)^2
+ 2  K_X \left(\frac{\pi'}{b} \right)^3 \l \frac{a'}{ba} + d \frac{c'}{bc} \r
\; .
\nonumber
\end{align}
We are interested in the combination
\be
\frac{1}{2} \l T_{0}^0 -  T_{r}^r \r
=  \left(\frac{\pi^\prime}{b} \right)^2
\left[F_X - K_\pi + K_X \frac{1}{b} \left(\frac{\pi^\prime}{b}\right)^\prime
- K_X \frac{\pi^\prime}{b} \left(\frac{a'}{ab} + d \frac{c^\prime}{cb}
\right)
\right] \; ,
\label{sep25-15-1}
\ee
which has to violate the ANECs of Sec.~\ref{ANEC}.

\subsection{Stability conditions}

We now turn to the discussion of the stability of Galileon perturbations
about static, spherically symmetric backgrounds $\pi_c(r)$, and write
 $\pi = \pi_c  + \chi$.
We are interested in high momentum and frequency modes, so we concentrate
on terms involving $\nabla_\mu \chi \nabla_\nu \chi$ in the
quadratic Largangian or, equivalently, second order terms,
proportional to
$\nabla_\mu \nabla_\nu \chi$, in the linearized field equation. 
A subtlety here is that the Galileon field equation involves the
second derivatives of metric, and the Einstein equations involve the
second derivatives of the Galileon~\cite{Deffayet:2010qz} (see also
ref.~\cite{Kobayashi:2010cm}), and so do the linearized equations 
for perturbations. The trick is to integrate the metric perturbations out
of the Galileon field equation by making use of the Einstein 
equations~\cite{Deffayet:2010qz}.

The full Galileon field equation reads
\begin{align*}
 \left( -2F_X + 2 K_\pi - 2K_{X \pi} \nabla_\mu \pi
\nabla^\mu \pi  - 
2 K_X \Box \pi \right) \Box \pi   +
\left(-4 F_{XX} + 4 K_{X \pi} \right) 
\nabla^\mu \pi \nabla^\nu \pi \nabla_\mu \nabla_\nu \pi &
\\
 - 4 K_{XX} \nabla^\mu \pi \nabla^\nu \pi \nabla_\mu \nabla_\nu \pi
\Box \pi  +
4 K_{XX} \nabla^\nu \pi \nabla^\lambda \pi \nabla_\mu \nabla_\nu \pi
\nabla^\mu \nabla_\lambda \pi + 2 K_X \nabla^\mu \nabla^\nu \pi 
\nabla_\mu \nabla_\nu \pi &
\\
 + 2 K_X R_{\mu \nu} \nabla^\mu \pi \nabla^\nu \pi + \ldots = 0 \; ; &
\end{align*}
hereafter dots denote terms without second derivatives.
The subtle term is the last one here. The linearized equation
can be written in the following form (hereafter we omit the
subscript $c$ in the notation for the Galileon background):
\begin{align}
 -2 [F_X  + K_X \Box \pi - K_\pi +  \nabla_\nu (K_X \nabla^\nu \pi)]
 \nabla_\mu \nabla^\mu \chi  &
\nonumber \\
-2  [2 (F_{XX} + K_{XX} \Box \pi) \nabla^\mu \pi \nabla^\nu \pi 
- 2(\nabla^\mu K_X) \nabla^\nu \pi - 2K_X \nabla^\mu \nabla^\nu \pi]
 \nabla_\mu \nabla_\nu \chi &
\nonumber\\
 + 2 K_X R_{\mu \nu}^{(1)} \nabla^\mu \pi \nabla^\nu \pi + \ldots &= 0 \; ,
\label{oct23-15-1} 
\end{align}
where the terms without the second derivatives of $\chi$ are omitted,
and $R_{\mu \nu}^{(1)}$ is linear in metric perturbations.
We now make use of the Einstein equations $R_{\mu \nu} - 
\frac{1}{2} g_{\mu \nu}R = T_{\mu \nu}$, or
\[
R_{\mu \nu} = T_{\mu \nu} - \frac{1}{d} g_{\mu \nu} T^\lambda_\lambda \; ,
\]
linearize the energy-momentum tensor and obtain for the last term 
in eq.~\eqref{oct23-15-1}
\be
 2 K_X R_{\mu \nu}^{(1)} \nabla^\mu \pi \nabla^\nu \pi
= - 2 K_X^2 \left[-\frac{2(d-1)}{d} X^2 \Box \chi + 4X
\nabla^\mu \pi \nabla^\nu \pi \nabla_\mu \nabla_\nu \chi \right] + \ldots \; .
\label{oct23-15-2}
\ee
The resulting linearized Galileon field equation is obtained from the
following quadratic Lagrangian:
 \begin{align}
L^{(2)} &= [F_X  + K_X \Box \pi - K_\pi +  \nabla_\nu (K_X \nabla^\nu \pi)]
 \nabla_\mu \chi \nabla^\mu \chi
\nonumber \\
&+ [2 (F_{XX} + K_{XX} \Box \pi) \nabla^\mu \pi \nabla^\nu \pi 
- 2(\nabla^\mu K_X) \nabla^\nu \pi - 2K_X \nabla^\mu \nabla^\nu \pi]
 \nabla_\mu \chi \nabla_\nu \chi
\nonumber\\
&+\delta L^{(2)} \; ,
\nonumber
\end{align}
where $\delta L^{(2)}$  corresponds to the term \eqref{oct23-15-2}:
\[
\delta L^{(2)} =  -\frac{2(d-1)}{d} K_X^2 X^2 \nabla_\mu \chi
\nabla^\mu \chi + 4K_X^2 X
\nabla^\mu \pi \nabla^\nu \pi \nabla_\mu \chi \nabla_\nu \chi \; .
\]
If we did not set $8\pi G =1$, the term $\delta L^{(2)}$ would contain
a factor $8\pi G$, while the rest of the quadratic Lagrangian would be
independent of $G$.

Specifying to static, spherically symmetric background, we find
\be
L^{(2)} = a^{-2} \tilde{\cal G}^{00} \dot{\chi}^2 - b^{-2} 
\tilde{\cal G}^{rr} (\chi')^2 
- c^{-2} \tilde{\cal G}^\Omega \gamma^{\alpha \beta}
\d_\alpha \chi \d_\beta \chi \; ,
\nonumber
\ee
where the effective metric is
\[
\tilde{\cal G}^{\mu \nu} = {\cal G}^{\mu \nu} + \delta  {\cal G}^{\mu \nu} \; ,
\]
with 
\begin{subequations}
\begin{align}
{\cal G}^{00} &=  F_X - K_\pi - \frac{K_X^\prime}{b} \frac{\pi'}{b}
- 2 K_X \frac{1}{b} \left(\frac{\pi'}{b}\right)^\prime - 
2 d K_X \frac{c^\prime}{cb}\frac{\pi'}{b} \; ,
\label{sep22-15-41} \\
{\cal G}^\Omega &=  F_X - K_\pi - \frac{K_X^\prime}{b} \frac{\pi'}{b}
- 2 K_X \frac{1}{b} \left(\frac{\pi'}{b}\right)^\prime - 
2 (d-1) K_X \frac{c^\prime}{cb}\frac{\pi'}{b}  - 
2  K_X \frac{a^\prime}{ab}\frac{\pi'}{b} \; ,
\label{sep22-15-42}\\
{\cal G}^{rr} &=  F_X - 2 F_{XX} \left(\frac{\pi^\prime}{b}\right)^2
- K_\pi 
+ \frac{K_X^\prime}{b} \frac{\pi^\prime}{b} - 2 K_X \frac{\pi^\prime}{b} 
\l \frac{a'}{ba} + d \frac{c'}{bc} \r
\nonumber \\
& + 2 K_{XX}  \left(\frac{\pi^\prime}{b}\right)^2 \frac{1}{b}  
\left(\frac{\pi^\prime}{b}\right)^\prime + 2 K_{XX}  
\left(\frac{\pi^\prime}{b}\right)^3 \l \frac{a'}{ba} + d \frac{c'}{bc} \r\; ,
\end{align}
\end{subequations}
with $K_X^\prime = d K_X/d r$ and
\begin{align*}
\delta {\cal G}^{00} & = \delta {\cal G}^{\Omega} = 
 -\frac{2(d-1)}{d} K_X^2  \left(\frac{\pi^{\prime}}{b}\right)^4 
\\
\delta {\cal G}^{rr} &= \frac{2(d+1)}{d} 
K_X^2  \left(\frac{\pi^{\prime}}{b}\right)^4 \; .
\end{align*}
The stability conditions for the 
perturbations are
\be
\tilde{\cal G}^{00} > 0 \; , \;\;\;\; 
\tilde {\cal G}^{rr} \geq 0  \; , \;\;\;\; \tilde {\cal G}^{\Omega} \geq 0 \; .
\label{sep22-15-40}
\ee
There are no ghosts and/or gradient instabilities only if
these conditions are satisfied.

Since both $\delta {\cal G}^{00}$ and $\delta {\cal G}^{\Omega}$ vanish
in the case of three-dimensional space-time ($d=1$) and are negative
in higher dimensions,
the first and third stability conditions in eq.~\eqref{sep22-15-40}
imply
\be
{\cal G}^{00} > 0 \; , \;\;\;\; 
{\cal G}^{\Omega} > 0 \; .
\label{oct23-15-3}
\ee
We will see that these necessary 
conditions, taken  together with
the ANEC violation, are difficult to satisfy.

We note in passing that at least in theories which have conventional
and stable
scalar field theory limit as $\d \pi \to 0$, namely, $F \to X - V(\pi)$
(and $K \to 0$, see eq.~\eqref{sep24-15-10}),
there is unavoidable superluminality near this limit. Indeed, by choosing
the background with $\dot{\pi} = \d_\alpha \pi =0$ at a given moment
of time and setting $a=b=1$, $c=r$ 
one can always have, with an appropriate sign of 
$\pi'$,
\[
\tilde{\cal G}^{\Omega} - \tilde{\cal G}^{00} = 2 K_X \frac{\pi'}{r} 
> 0 \; ,
\]
which means that the modes normal to the radial direction propagate
superluminally. At small $\pi'$ such a background is stable, since
in this regime $\tilde{\cal G}^{00} = \tilde{\cal G}^{\Omega} = 
\tilde{\cal G}^{rr} = F_X$ modulo small corrections.
This observation is in line with earlier results on the  superluminality
of Galileons~\cite{superlum2}.

\section{Tensions, No-Go's}
\label{Tensions}

\subsection{Generalities}

Let us see that the necessary conditions for
the stability, eq.~\eqref{oct23-15-3}, are in tension
with the ANEC violation discussed in Sec.~\ref{ANEC}. To this end, we 
multiply ${\cal G}^{00}$ given by eq.~\eqref{sep22-15-41} by 
$\mu \cdot (\pi'/b)^2$, where $\mu (r)$ is yet unspecified positive
function, 
integrate over $r$ from $-\infty$ to $+\infty$,
and integrate by parts 
the third term in the right hand side of eq.~\eqref{sep22-15-41}, setting
(recall our assumption, eq.~\eqref{sep23-15-2})
\be
 \pi^{\prime 3}  K_X \mu  \to 0 \;\;\;\;\;
\mbox{as} \;\; r \to \pm \infty \; .
\label{sep22-15-55}
\ee
We obtain
\begin{align}
&\int_{-\infty}^{+\infty}~dr~\mu(r) \l \frac{\pi'}{b} \r^2 {\cal G}^{00} 
\nonumber\\
 =
&\int_{-\infty}^{+\infty}~dr~\mu(r) \l \frac{\pi'}{b} \r^2 \left[ F_X - K_\pi 
+ K_X \frac{1}{b} \left(\frac{\pi'}{b}\right)^\prime +
 K_X \frac{\pi'}{b} \frac{1}{b} \l \frac{\mu'}{\mu} - \frac{b'}{b} -2d
\frac{c'}{c} \r \right] > 0 \; .
\label{sep22-15-50}
\end{align}
We now choose 
\[
\mu = \frac{bc^d}{a} 
\]
and see that the integrand in the right hand side of eq.~\eqref{sep22-15-50}
is proportional to $(T^0_0 - T^r_r)$ given by eq.~\eqref{sep25-15-1}. Thus,
\be
\int_{-\infty}^{+\infty}~dr~\frac{bc^d}{a} \l T^0_0 - T^r_r \r > 0 \; .
\label{sep22-15-60}
\ee
This is in tension with eq.~\eqref{sep22-15-45}, although
for $d \geq 2$ there is no direct contradiction.

By performing the same procedure with ${\cal G}^\Omega$ with the measure
\[
\mu = abc^{d-2}
\]
we obtain
\be
\int_{-\infty}^{+\infty}~dr~abc^{d-2} \l T^0_0 - T^r_r \r > 0 \; .
\label{sep24-15-1}
\ee
More generally, we consider a combination
\[
(1-\beta) {\cal G}^{00} + \beta {\cal G}^\Omega > 0 \;\;\;\;\;\; 0\leq \beta \leq 1 \; ,
\]
choose the measure as
\[
\mu = a^{2\beta -1} b c^{d-2\beta} 
\]
and find
\be
\int_{-\infty}^{+\infty}~dr~a^{2\beta -1} b c^{d-2\beta} \l T^0_0 - T^r_r \r > 0 
 \;\;\;\;\;\; \mbox{for~all} \; \;  0\leq \beta \leq 1
\; ,
\label{sep22-15-63}
\ee
which should hold together with the ANEC violation inequality 
\eqref{sep22-15-45}. Obviously, there is tension 
between these inequalities. As an example, for $d=2$ (4-dimensional
space-time) one can choose $\alpha =1$ in eq.~\eqref{sep22-15-45}
and $\beta = 1/2$ in eq.~\eqref{sep22-15-63} to get
\begin{align}
&\int_{-\infty}^{+\infty}~dr~ \frac{bc}{a} \l T^0_0 - T^r_r \r < 0 \; ,
\nonumber\\
&\int_{-\infty}^{+\infty}~dr~ bc \l T^0_0 - T^r_r \r > 0 \; .
\nonumber 
\end{align}
This shows that the Galileon-supported wormholes, if any, must be 
quite tricky.

We note that our assumption \eqref{sep22-15-55} is very mild.
The large-$|r|$ behavior of the relevant measures is at most $|r|^d$,
so we need $\pi^\prime = o (|r|^{-d/3})$. On the other hand,
if the Galileon becomes an ordinary
scalar field in the weak-field limit, one has $\pi^\prime \propto |r|^{-d}$,
which is more than sufficient.
In other words, to violate the
assumption \eqref{sep22-15-55} and at the same time have the Minkowski
limit at $\d \pi =0$, the Galileon must be pretty contrived.

\subsection{3-dimensional space-time}

In the case of 3-dimensional space-time we have $d=1$, so there is direct 
contradiction between the inequalities \eqref{sep22-15-45}  with $\alpha=1$
and
\eqref{sep22-15-60}. 
So, in this case we have a no-go theorem
stating that there are no stable, static, spherically symmetric
wormholes in Galileon theories with the Lagrangians of the form 
\eqref{sep22-15-30} and Minkowski limit at $\d \pi =0$.
 The only way to get around this theorem
is to violate the property \eqref{sep22-15-55}, which reads
\[
r \pi^{\prime 3} K_X  \to 0  \;\;\;\;\;
\mbox{as} \;\; r \to \pm \infty \; .
\]
Attempting to explore this loophole is not promising, we think.

\subsection{Space-times of more than 3 dimensions}

For $d \geq 2$, including the most physically interesting case of 
4-dimensional space-time ($d=2$), there is no direct contradiction
between the ANEC violation inequality   \eqref{sep22-15-45} and stability
inequality \eqref{sep22-15-63}. Yet  the possible shapes of  wormholes
are strongly constrained. To this end, we first
consider
eqs.~\eqref{sep23-15-1} and \eqref{sep24-15-1}. Taken together,
these rule out wormholes
for which
\[
    \vert \frac{dc}{dR} \vert \leq 1 \;\;\;\;\; \mbox{for~all} \;\; r \; ,
\] 
including wormholes with monotonous $dc(R)/dR$, Fig.~\ref{fig2}. 

Another
constraint follows directly from eq.~\eqref{sep22-15-15}. Namely, consider
the often used (``Schwarzschild'') coordinate frame, in which
\[
   b(r) = a^{-1} (r) \; .
\]
The function $c'(r)$ 
cannot be  monotonous in this frame either.
Indeed, 
if $c'$ 
monotonously increases in this frame
from $-1$ to $1$ as $r$ runs from
$-\infty$ to $+\infty$, then $c'' > 0$, and
$(T^0_0 - T^r_r) < 0$ everywhere. This contradicts any of the inequalities
\eqref{sep22-15-63}. 

These two properties rule out the simplest 
wormhole shapes.

To conclude this Section we note that adding conventional matter that does not
violate the NEC would not help. The ANEC violation inequalities of 
Sec.~\ref{ANEC} must hold for the total energy-momentum tensor, and since the
conventional matter has $T^0_0 - T^r_r > 0$, these inequalities must still
be valid for the Galileon contribution to the total $T^\mu_\nu$. Thus,
our analysis remains intact.

\section{Discussion}
\label{Discussion}

Even though our findings are not completely conclusive, they show
that constructing Galileon-supported wormholes must be tricky, if at
all possible. One way to get around of our constraints would be to give
up our initial assumption that the Minkowski regime occurs at 
$\d \pi = 0$, which lead to eq.~\eqref{sep23-15-2}. This is indeed a
possibility if $F=F(X)$, $K=K(X)$ depend on $X=(\d \pi)^2$ 
but not on $\pi$ itself.
Then the 
linear Galileon background, say, $\pi = Q x^1$ where $Q$ is a constant,
obeys the Galileon field equation and has vanishing energy-momentum tensor
provided that $F(-Q^2) = 0$ and $F_X (-Q^2) =0$, somewhat
resembling the ghost condensate case~\cite{ArkaniHamed:2003uy}. 
In our spherically symmetric setting, the Galileon with asymptotics
$\pi \to \pm Qr $ as $r \to \pm \infty$ would violate eq.~\eqref{sep22-15-55},
so our arguments would not work. Unlike in the ghost condensate
case, however, the dispersion relation for perturbations about
$\pi = Qx^1$ at quadratic level in momenta and frequency is $(p^1)^2 = 0$.
As discussed in Ref.~\cite{Dubovsky:2004sg}, this is problematic from
the effective field theory viewpoint: genuine higher order terms
would modify the dispersion relation to
\[
(p^1)^2 = A \frac{\omega^4}{\Lambda^2} \; ,
\]
where $\Lambda$ is a UV cutoff, and $A$ is generically of order one. 
This would yield the gradient
instability with the time scale as small as
$\Lambda^{-1}$. Nevertheless, one can assume 
that $A$ is fine tuned to be very small, so searching for stable
wormholes
with $\pi \to \pm Qr $ at large $|r|$ is of interest.

The author is indebted to M. Libanov for helpful discussions
and to F.~Canfora and S.~Deser for useful correspondence.
Special thanks are to A.~Vikman who pointed out a drawback
in the original analysis.
This work has been supported by Russian Science Foundation
grant 44-22-00161.

\end{document}